\documentclass{JHEP3}
\usepackage{epsfig}
\usepackage{dcolumn}
\usepackage{bm}
\newcommand{\be}{\begin{equation}}
\newcommand{\ee}{\end{equation}}
\def\bea{\begin{eqnarray}}
\def\eea{\end{eqnarray}}

\usepackage{amssymb}
\usepackage{amsmath}
\usepackage{amsfonts}
\usepackage{graphicx}
\usepackage{psfrag}

 \def\be{\begin{equation}}
\def\ee{\end{equation}}
\def\bea{\begin{eqnarray}}
\def\eea{\end{eqnarray}}
\def\lesssim{\mathrel{\hbox{\rlap{\hbox{\lower4pt\hbox{$\sim$}}}\hbox{$<$}}}}
\def\gtrsim{\mathrel{\hbox{\rlap{\hbox{\lower4pt\hbox{$\sim$}}}\hbox{$>$}}}}
\def\jbar{\bar{j}}
\title{Features of deSitter Vacua in M-Theory }
\author{ Jonathan P. Hsu$^a$\footnote{\mbox{
Email: {\tt pihsu@stanford.edu} }}, Renata
Kallosh$^a$\footnote{\mbox{ Email: {\tt kallosh@stanford.edu} }} and
Navin Sivanandam$^{a,b}$\footnote{\mbox{ Email: {\tt
navins@stanford.edu} }}
\\
$^a$Department of Physics, Stanford University, Stanford, CA 94305,
USA \\
$^b$SLAC, Stanford University, Stanford, CA 94309, USA}
\received{\today} 
 \preprint{SU-ITP-05/31 \\
  SLAC-PUB-11548\\ October 28, 2005}
\abstract{ We compute the masses of all moduli in the unstable
deSitter vacua arising in the toy model of cosmological M-theory
flux compactifications on the $G_2$ holonomy manifolds of
\cite{acharya1}. The slow-roll parameters in the tachyonic
directions are shown to be too large to be useful for conventional
models of inflation. However, it appears that we can find fast roll
regimes which could, under certain conditions, account for the
current dark energy driven accelerated expansion of the universe.}
\begin{document}

\section{Introduction}

There is a long history of studying deSitter vacua in string theory.
Early development included the study of \cite{Kallosh:2001gr} which
identified deSitter space as a solution in the gauged extended
supergravities that arise as consistent truncations of
10-dimensional type II supergravity and 11-dimensional supergravity
in backgrounds of the form $dS_4\times H^{p,q}$ where the second
factor is a hyperbolic space (of six or seven dimensions). These
backgrounds are noncompact and hence the gauging is a noncompact
one. Other studies of these ghost-free supergravities included
\cite{gibbons_cvetic}, \cite{gibbons_hull}.

The study of deSitter space in a stable compactification of type II
string theory was initiated by KKLT \cite{kklt}. A mechanism of
perturbative stabilization of M-theory vacua was proposed in
\cite{acharya_modfix} by turning on flux on the internal space as
well as turning on flux on singular loci within the internal space.
In another line of investigation, moduli stabilization was achieved
by non-perturbative contributions to the superpotential following
work on determining more precise expressions for the K\"{a}hler
potential in compactifications on $G_2$ manifolds realized as
blowups of singular seven tori \cite{lukasmorris1}
\cite{lukasmorris2} \cite{lukasmorris3}.

In \cite{acharya1} the authors were interested in the statistics of
vacua of M-theory on manifolds of $G_2$ holonomy. They used a toy
model which is similar to the one discussed in
\cite{acharya_modfix}, where a perturbative Chern-Simons
contribution is necessary to stabilize the moduli. The model shares
the feature of the pre-KKLT models in that its deSitter vacua have
tachyonic directions. We will attempt to take this model more
seriously by computing the cosmological parameters that they
produce.

It is widely accepted that the universe (or more precisely our
Hubble volume) has undergone two periods of accelerated expansion in
its history. The first of these is inflation \cite{Linde:Book},
which took place in some earlier epoch, and the second is the dark
energy driven expansion that we observe today (see \cite{Riess:2004}
for recent data). Both of these accelerated periods can result from
slowly-rolling scalar fields. Observational constraints can be
placed on the parameters $\varepsilon$, $\eta$ that quantify the
degree of slow roll allowing us to assess the usefulness of this
model for cosmological purposes.

The scalar field potential we will obtain actually gives a somewhat
more involved model than is often used. The deSitter vacuum and
associated tachyonic directions will give accelerated expansion.
However, such expansion will only continue until the potential runs
negative, resulting in the collapse of the universe. In this case
our concern (for dark energy at least) is that the potential is flat
enough and that the field starts close enough to the maximum so as
to be consistent with the age of the observable universe. Such
models are discussed in detail in \cite{Kallosh:2001gr},
\cite{Kallosh:2002gg} and \cite{Linde:2001}.

In another interesting direction, it was noted in
\cite{Kallosh:2002wj} that the examples of deSitter space from
gauged supergravity truncations of type II and 11-dimensional
supergravity had the strange property that the scalar mass spectrum
was quantized in units of $1/3$ of the cosmological constant. It was
shown that the mass of a scalar was a particular Casimir in the
algebra of isometries of $dS_4$ but there was no reason for that
Casimir to be quantized. The list of examples of deSitter vacua with
this behavior was quite impressive, including all examples from
gauged $\mathcal{N}=2,4$ and 8 supergravity studied in
\cite{Kallosh:2001gr}. The feature was found to persist in models
with tachyon-free dS vacua \cite{fre_trigiante_vanproeyen}. It turns
out that the M-theory models of \cite{acharya1}, with appropriate
fluxes, generate scalar masses that may or may not have quantized
scalars.

\section{The Model}

We are interested in a compactification of M-theory on a manifold
$X$ of $G_2$ holonomy. We use the conventions of \cite{acharya1}
which we will summarize. The resulting theory in four dimensions is
an $\mathcal{N} =1$ supergravity specified by the field content,
K\"{a}hler potential and superpotential. The fields are, as usual,
the complexified coordinates on the moduli space, $\mathcal{M}$, of
metric deformations on $X$. This moduli space has dimension
$n=b^3(X)$ and has complexified coordinates given by the periods of
the $G_2$ invariant three-form $\Phi$ complexified by the periods of
the three-form potential $C_3$,
\begin{equation}
z^i = t^i + i s^i = \frac{1}{l_M^3}\int(C_3 + i\Phi) \,\, .
\end{equation}
The moduli space of metrics is K\"{a}hler with K\"{a}hler potential given by
\begin{equation}
K(z,\bar{z})=-3\ln(4\pi^{\frac{1}{3}}V_X(s)) \,\, .
\end{equation}
There are no known strong constraints on the dependence of the volume of the
manifold $X$ on the moduli $s$ except that it be a homogeneous
function of degree $7/3$ and that the above constructed $K$ is convex.
Thus the most general function $V_X(s)$ is
\begin{equation}
V_X(s)=\prod_{k=1}^n s_k^{a_k}f(s)
\end{equation}
with the exponents $a_k$ such that
\begin{equation}
\sum_{k=1}^n a_k=\frac{7}{3} \,\, .
\end{equation}
The arbitrary function $f(s)$ is invariant under rescaling and thus, to
leading order, can be taken to be one
\begin{equation}\label{volume}
V_X(s)=\prod_{k=1}^n s_k^{a_k} \,\, .
\end{equation}
As usual, we turn on an internal flux of $G_4$ which induces a GVW
superpotential \cite{gvw}
\begin{equation}
W \sim \int_X (C_3 + i \Phi)\wedge G_4 \,\, .
\end{equation}
As described in \cite{acharya_modfix}, we'll actually need more than this
to stabilize the moduli. Consider $G_2$ manifolds $X$
that have an $ADE$ orbifold singularity. The center of the $ADE$ space is
a singular (and supersymmetric) three-cycle, $Q$. There are light M2
branes
localized at the singular three-cycle due to the
shrinking two-cycles of the $ADE$ space. These light degrees of freedom
combine to make a Chern-Simons gauge theory on $Q$ with gauge group
the complexification of the $ADE$ gauge group, $G^{\mathbb{C}}$. The
Chern-Simons background is determined by a flat connection which makes a
complex constant contribution to the superpotential. The final
superpotential is thus given by
\begin{equation}
W(z)=\frac{1}{\kappa_4^3}(N_i z^i + c_1 + i c_2) \,\, .
\end{equation}
The constant contribution $c_2$ will turn out to be a crucial ingredient.
Although there are no known explicit examples of how to compute these
constants, there is no good reason to forbid their existence.

\section{deSitter critical points}

Given the above expressions for the K\"{a}hler and superpotential,
the potential can be computed as usual,
\begin{equation}
V(z,\bar{z}) = \kappa_4^2 e^K (g^{i\jbar}D_iW D_{\jbar}\bar{W} - 3 |W|^2)
\end{equation}
which yields the following \cite{acharya1}:
\begin{equation} \label{potential}
V=\frac{c_2^2}{48\pi
V_X^3}\left(3+\sum_{j=1}^na_j\nu_js_j(\nu_js_j-3)\right)+\frac{1}{48\pi
V_X^3}\left(\overrightarrow{N}\cdot\overrightarrow{t}+c_1\right)^2
\end{equation}
where $\nu_i\equiv-\frac{N_j}{c_2a_j}$. In the single modulus case
(dropping the axion term), we have:
\begin{equation}
V=\frac{1}{16\pi}\left(\frac{c_2^2}{s^7}+\frac{Nc_2}{s^6}+\frac{N^2}{7s^5}\right)
\end{equation}
In order to calculate slow roll parameters we need to use
canonically-normalized scalar fields. The K\"{a}hler potential in
the previous section leads to the following K\"{a}hler metric:
\begin{equation}
\mathrm{d}s^2=\sum^n_{i=1}\frac{3a_i}{4s_i^2}\mathrm{d}z_i\mathrm{d}\bar{z}_i=\sum_{i=1}^n\left(\mathrm{d}t_i^2+\mathrm{d}s_i^2\right)
\end{equation}
This will give a kinetic term that looks like:
\begin{equation}
g_{i\bar{j}}\partial_{\mu}z^i\partial^{\mu}\bar{z}^{\bar{j}}=\sum_i\frac{3a_i}{4s_i^2}\left(\partial_{\mu}t_i\partial^{\mu}t_i+\partial_{\mu}s_i\partial^{\mu}s_i\right)
\end{equation}
Ignoring the $t$-terms we see that if we define:
\begin{equation}
s_i=\exp\left(\sqrt{\frac{2}{3a_i}}\phi_i\right)
\end{equation}
Then, $\phi_i$ are the canonically normalized moduli fields. For our
single modulus case we will have
$s=\exp\left(\sqrt{\frac{2}{7}}\phi\right)$, which then gives the
potential as:
\begin{equation}
V(\phi)=\frac{1}{16\pi}\left(c_2^2e^{\left(-7\sqrt{\frac{2}{7}}\right)\phi}
+Nc_2e^{\left(-6\sqrt{\frac{2}{7}}\right)\phi}
+\frac{N^2}{7}e^{\left(-5\sqrt{\frac{2}{7}}\right)\phi}\right)
\end{equation}
We then straightforwardly obtain:
\begin{eqnarray}
V'(\phi)&=&-\frac{1}{16\pi}\sqrt{\frac{2}{7}}\left(7c_2^2e^{\left(-7\sqrt{\frac{2}{7}}\right)\phi}
+6Nc_2e^{\left(-6\sqrt{\frac{2}{7}}\right)\phi}
+\frac{5N^2}{7}e^{\left(-5\sqrt{\frac{2}{7}}\right)\phi}\right)\\
V''(\phi)&=&\frac{1}{56\pi}\left(49c_2^2e^{\left(-7\sqrt{\frac{2}{7}}\right)\phi}
+36Nc_2e^{\left(-6\sqrt{\frac{2}{7}}\right)\phi}
+\frac{25N^2}{7}e^{\left(-5\sqrt{\frac{2}{7}}\right)\phi}\right)
\end{eqnarray}
We can set $V'=0$ and find the critical points. The unstable de
Sitter maxima corresponds to $s=-\frac{7c_2}{N}$. The slow roll
parameters are defined as:
\begin{eqnarray}
\varepsilon&=&\frac{1}{2}\left(\frac{V'}{V}\right)^2\\
\eta&=&\frac{V''}{V}
\end{eqnarray}
To compute these parameters around the critical point,
$\tilde{\phi}=\sqrt{\frac{7}{2}}\ln\left(\frac{-7c_2}{N}\right)$, we
set $\phi=\tilde{\phi}+\delta$ and eventually obtain:
\begin{eqnarray}
\varepsilon&=&\frac
{28\left(-3+3\mathrm{cosh}\left[\sqrt{\frac{2}{7}}\delta\right]
+2\mathrm{sinh}\left[\sqrt{\frac{2}{7}}\delta\right]\right)^2}
{\left(-7+8\mathrm{cosh}\left[\sqrt{\frac{2}{7}}\delta\right]
+6\mathrm{sinh}\left[\sqrt{\frac{2}{7}}\delta\right]\right)^2}\\
\eta&=&\frac{50}{7}+\frac{48-154e^{\sqrt{\frac{2}{7}}\delta}}
{7-49e^{\sqrt{\frac{2}{7}}\delta}+49e^{2\sqrt{\frac{2}{7}}\delta}}
\end{eqnarray}
Expanding around $\delta=0$ gives:
\begin{eqnarray}
\varepsilon&=&32\delta^2-48\sqrt{14}\delta^3+O[\delta^4]\\
\eta&=&-8+12\sqrt{14}\delta-\frac{880}{7}\delta^2+\frac{4636}{7}\sqrt{\frac{2}{7}}\delta^3+O[\delta^4]
\end{eqnarray}
These are neither small, nor obviously tunable -- since they are
independent of fluxes.

Using the results and machinery of \cite{acharya1} we can generalise
this to n moduli fields. First, note that with the
canonically-normalized $s_i=\exp^{\sqrt{\frac{2}{3a_i}}\phi_i}$ and
our earlier expression for $V$ (\ref{potential}), we will get:
\begin{equation}
\frac{\partial
V}{\partial\phi_i}=\sqrt{\frac{2a_i}{3}}\frac{c_2^2}{48\pi
V_X^3}\left(-3E+2\nu_i^2s_i^2-3\nu_is_i\right)
\end{equation}
\begin{equation}
\frac{\partial^2V}{\partial\phi_j\partial\phi_i}=\frac{2}{3}\frac{c_2^2}{48\pi
V_X^3}\left(3\sqrt{a_ia_j}
\left(3E-2\left(\nu_i^2s_i^2+\nu_j^2s_j^2\right)+3\left(\nu_is_i+\nu_js_j\right)\right)
+\delta{ij}\left(4\nu_i^2s_i^2-3\nu_is_i\right)\right)
\end{equation}
$E$ is defined as
\begin{equation}
E=3+\sum_{j=1}^n a_j\nu_js_j\left(\nu_js_j-3\right)
\end{equation}
The critical points are obtained in \cite{acharya1}. If we define
$h_i=\nu_is_i$, then the solutions are:
\begin{equation}
h_i=\frac{3}{4}+m_iH
\end{equation}
With $m_i=\pm1$, $A=\vec{a}\cdot\vec{m}$ and $H$ given by:
\begin{equation}
H(A)=\frac{3}{20}\left(3A-\sqrt{9A^2+15}\right)
\end{equation}

We'll rewrite the $\eta$ matrix at a critical point determined by the
$m_i$ as follows
\begin{equation}
\eta_{ij}(a_i,m_i) = Q(A) \sqrt{a_i a_j} +S(A) \delta_{ij} +
T(A)\left(\frac{1-m_j}{2} \right) \delta_{ij}
\end{equation}
where
\begin{equation}
\begin{array}{rcl}
Q(A) & = & \frac{2}{E(A)}(3E(A) +9/4 -4H(A)^2) \\
S(A) & = & \frac{2H(A)}{E(A)}\left( \frac{4H(A)}{3} +1 \right) \\
T(A) & = & \frac{-4H(A)}{E(A)} \\
E(A) & = & \frac{-15}{16} - \frac{3}{2} A H(A) + \frac{7}{3}H(A)^2 \\
\end{array}
\end{equation}

As pointed out by Acharya et al., the physical vacua can be put into three
categories determined by requiring the moduli fields to be positive.
When all
fluxes are negative, there are several $AdS$ minima with $A > -1/3$ as well
as
one $dS$ maximum with $A=-7/3$ when all $m_i=-1$. When some of the fluxes
are not negative and $A <-1/3$ then there is a single physical vacuum which is a $dS$
maximum when $sign(N_i) = sign(m_i)$. When some of
the fluxes are negative and $A > -1/3$ then there are
no vacua that have positive moduli. We will now consider the
slow roll parameters of the $dS$ vacua in the two cases when they occur.

\subsection{All negative fluxes}

For negative fluxes, the $dS$ vacuum is at all $m_i=-1$. This gives
for $\eta$
\begin{equation}
\eta_{ij}(a_i, m_i=-1)=V^{-1} \frac{\partial^2 V}{\partial \phi_i \partial
\phi_j} = 6(\delta_{ij} - \sqrt{a_i a_j}) \,\,\, .
\end{equation}
We are interested in eigenvalues of this matrix. Define $\alpha_i =
\sqrt{a_i}$.
Note that $\alpha_i$ is a n-vector with norm $\sum_{i=1}^n \alpha_i
\alpha_i = 7/3$.
Then the matrix is
\begin{equation}
\eta_{ij} =  6(\delta_{ij} - \alpha_i \alpha_j)
\end{equation}
So first, it is clear that $\alpha_i$ is an eigenvector of $\eta$ with
eigenvalue -8. Since $\alpha_i$ is a nonzero n-vector, there are n-1
orthogonal vectors. Each such orthogonal vector is in fact an eigenvector
of $\eta$ with eigenvalue 6. Hence the eigenvalues of $\eta$ are -8 and 6
with multiplicity 1 and n-1. Needless to say, this is independent of what
values the $a_i$ take as long as they satisfy the constraint $\sum_i a_i
= 7/3$.

It was discussed in \cite{Kallosh:2002wj} that the Casimir operator in $dS_4$ space,
$k={3 m^2\over \Lambda}= {m^2\over H^2}=3\eta$, in all known extended
supergravities takes integer values. There is no known explanation of this
observation. Here we find again that $3m^2/\Lambda$ = -24, 18 are integral.
\subsection{Some negative fluxes}
To make some progress in this more general case let us consider
again the form of the $\eta$ matrix:
\begin{equation}
\eta_{ij}(a_i,m_i) = Q(A) \sqrt{a_i a_j} +S(A) \delta_{ij} +
T(A)\left(\frac{1-m_j}{2} \right) \delta_{ij}
\end{equation}
where
\begin{equation}
\begin{array}{rcl}
Q(A) & = & \frac{2}{E(A)}(3E(A) +9/4 -4H(A)^2) \\
S(A) & = & \frac{2H(A)}{E(A)}\left( \frac{4H(A)}{3} +1 \right) \\
T(A) & = & \frac{-4H(A)}{E(A)} \\
E(A) & = & \frac{-15}{16} - \frac{3}{2} A H(A) + \frac{7}{3}H(A)^2 \\
\end{array}
\end{equation}
Let's consider the case when we have $k$ negative fluxes and the
rest positive. Since we are free to reorder the $a_i$ we can,
without loss of generality, set $N_1$ through to $N_k$ to be
negative, while $N_{k+1}$ through $N_n$ are positive. This then
gives:
\begin{equation}
m_i=sign(N_i)=\left\{\begin{array}{l}
-1\textrm{ if $i=1\ldots k$}\\
1\textrm{ if $i=k+1\ldots n$}\\ \end{array}\right.
\end{equation}
With this in mind we then write $\eta$ as:
\begin{equation}
\eta=\left(\begin{array}{cc} M^-&q\\
q^T&M^+\end{array}\right)
\end{equation}
where
\begin{eqnarray}
M^-&=&Q(A)\sqrt{a_ia_j}+(S(A)+T(A))\delta_{ij}\qquad(i,j=1,\ldots,k)\\
M^+&=&Q(A)\sqrt{a_ia_j}+S(A)\delta_{ij}\qquad(i,j=k+1,\ldots,n) \\
q&=&Q(A)\sqrt{a_ia_j}\qquad(i=1,\ldots,k;j=k+1,\ldots,n)
\end{eqnarray}
Now we set $\alpha_i=\sqrt{a_i}$ and define the following:
\begin{eqnarray}
v^-_i&=&\left\{\begin{array}{l}
\alpha_i\textrm{ if $i=1\ldots k$}\\
0\textrm{ if $i=k+1\ldots n$}\\ \end{array}\right.\\
v^+_i&=&\left\{\begin{array}{l}
0\textrm{ if $i=1\ldots k$}\\
\alpha_i\textrm{ if $i=k+1\ldots n$}\\ \end{array}\right.\\
w^-_i&=&\left\{\begin{array}{l}
w_i\textrm{ if $i=1\ldots k$}\\
0\textrm{ if $i=k+1\ldots n$}\\ \end{array}\right.\\
w^+_i&=&\left\{\begin{array}{l}
0\textrm{ if $i=1\ldots k$}\\
w_i\textrm{ if $i=k+1\ldots n$}\\ \end{array}\right.
\end{eqnarray}
We choose $w^{\pm}$ such that $\alpha\cdot w^{\pm}=0$. This gives us
$k-1$ choices for $w^-$ and $n-k-1$ choices for $w^+$, a point we
shall return to momentarily. It is reasonably straightforward to
show that $w^{\pm}$ are eigenvectors of $\eta$:
\begin{eqnarray}
\eta w^-&=&\left(\begin{array}{cc} M^-&q\\
q^T&M^+\end{array}\right)\left(\begin{array}{c}w_i^-\\0\end{array}\right) \nonumber\\
&=&\left(\begin{array}{c}M^-_{ij}w_j^-\\q^T_{ij}w_j^-\end{array}\right) \nonumber\\
&=&\left(\begin{array}{c}\\\left(Q(A)\alpha_i\alpha_j+\left(S(A)+T(A)\right)\delta_{ij}\right)w_j^-\\Q(A)\alpha_i\alpha_jw_j^-\end{array}\right) \nonumber\\
&=&\left(\begin{array}{c}\left(S(A)+T(A)\right)w_i^-\\0\end{array}\right)\nonumber\\
&=&(S(A)+T(A))w^-
\end{eqnarray}
In the final step we used that fact that $\alpha\cdot w^-=0$ and
that $w_i$ is equal to zero for $i=k+1\ldots n$. Thus the $k-1$
$w^-$ are eigenvectors with $S(A)+T(A)$ the associated eigenvalue of
multiplicity $k-1$. An analogous argument follows for $w^+$, this
time giving an eigenvalue of $S(A)$ with a multiplicity of $n-k-1$.

We have found all but two of our eigenvalues. Sadly, though, things
are not quite so straightforward for the final two. Let us consider
the action of $\eta$ on $v^{\pm}$:
\begin{eqnarray}
\eta v^-&=&\left(\begin{array}{cc} M^-&q\\
q^T&M^+\end{array}\right)\left(\begin{array}{c}v_i^-\\0\end{array}\right)\nonumber\\
&=&\left(\begin{array}{c}M^-_{ij}v_j^-\\q^T_{ij}v_j^-\end{array}\right) \nonumber\\
&=&\left(\begin{array}{c}\\\left(Q(A)\alpha_i\alpha_j+\left(S(A)+T(A)\right)\delta_{ij}\right)\alpha_j\\Q(A)\alpha_i\alpha_j\alpha_j\end{array}\right) \nonumber\\
&=&\left(\begin{array}{c}\\\left(\left|\alpha_k^+\right|^2Q(A)+\left(S(A)+T(A)\right)\right)\alpha_i\\\left|\alpha_k^+\right|^2Q(A)\alpha_i\end{array}\right) \nonumber\\
&=&\left|\alpha_k^-\right|^2Q(A)(v^-+v^+)+(S(A)+T(A))v^-\qquad\left|\alpha_k^-\right|^2=\alpha_1^2+\ldots+\alpha_k^2
\end{eqnarray}
And, in a similar fashion, we obtain:
\begin{equation}
\eta
v^+=\left|\alpha_k^+\right|^2Q(A)(v^-+v^+)+S(A)v^+\qquad\left|\alpha_k^+\right|^2=\alpha_{k+1}^2+\ldots+\alpha_n^2
\end{equation}
Since $v^{\pm}$ are sent to linear combinations of $v^{\pm}$ by he
action of $\eta$ we know that the remaining two eigenvectors are
themselves linear combinations of $v^{\pm}$. Thus we can find the
remaining eigenvalues by diagonalizing the matrix:
\begin{equation}
\left(\begin{array}{cc}\left|\alpha_k^-\right|^2Q(A)+S(A)+T(A)&\left|\alpha_k^+\right|^2Q(A)\\
\left|\alpha_k^-\right|^2Q(A)&\left|\alpha_k^+\right|^2Q(A)+S(A)\end{array}\right)
\end{equation}
These eigenvalues are:
\begin{equation}
\frac{1}{2}\left(2S+T+Q\left(\left|\alpha_k^-\right|^2+\left|\alpha_k^+\right|^2\right)\pm\sqrt{T^2+2QT\left(\left|\alpha_k^-\right|^2-\left|\alpha_k^+\right|^2\right)+Q^2\left(\left|\alpha_k^-\right|^2-\left|\alpha_k^+\right|^2\right)^2}\right)
\end{equation}
However we also know that the constraints on the $a_i$ and the
definition of $A$ mean that:
\begin{equation}
\begin{array}{ccccc}
\left|\alpha_k^-\right|^2+\left|\alpha_k^+\right|^2&=&\sum_{i=1}^na_i&=&\frac{7}{3}\\
\left|\alpha_k^-\right|^2-\left|\alpha_k^+\right|^2&=&-\sum_{i=1}^na_im_i&=&-A
\end{array}
\end{equation}
Thus we get the final two eigenvalues as:
\begin{equation}
\lambda^{\pm}=\frac{1}{2}\left(\frac{7}{3}Q+2S+T\pm\sqrt{\frac{49}{9}Q^2-2AQT+T^2}\right)
\end{equation}
Although the final expressions for the eigenvalues of the $\eta$
matrix are somewhat unwieldy when expressed in terms of $A$ it can
be shown that they are monotonic functions in the range
$-\frac{7}{3}$ to $-\frac{1}{3}$. Plots of these eigenvalues can be
seen in Figure \ref{fig:eig}. The third graph -- corresponding to
the $\lambda^-$ eigenvalue -- is the only one which gives negative
eigenvalues. It is clear from reading off the boundary values that
the magnitude of the eigenvalue is bounded by $4$. Note,
however, that this eigenvalue is undefined at $A=-\frac{1}{3}$ so we
cannot saturate the bound.

\psfrag{a}{(a)} \psfrag{b}{(b)} \psfrag{c}{(c)} \psfrag{d}{(d)}
\psfrag{S}[cc][cc][1][90]{S} \psfrag{ST}[cr][cc][1][90]{S+T}
\psfrag{lminus}[cc][cc][1][90]{$\lambda^-$}
\psfrag{lplus}[cc][cc][1][90]{$\lambda^+$} \psfrag{A}{A}
\FIGURE[b,t]{
\includegraphics[width=\textwidth,height=0.5\textheight]{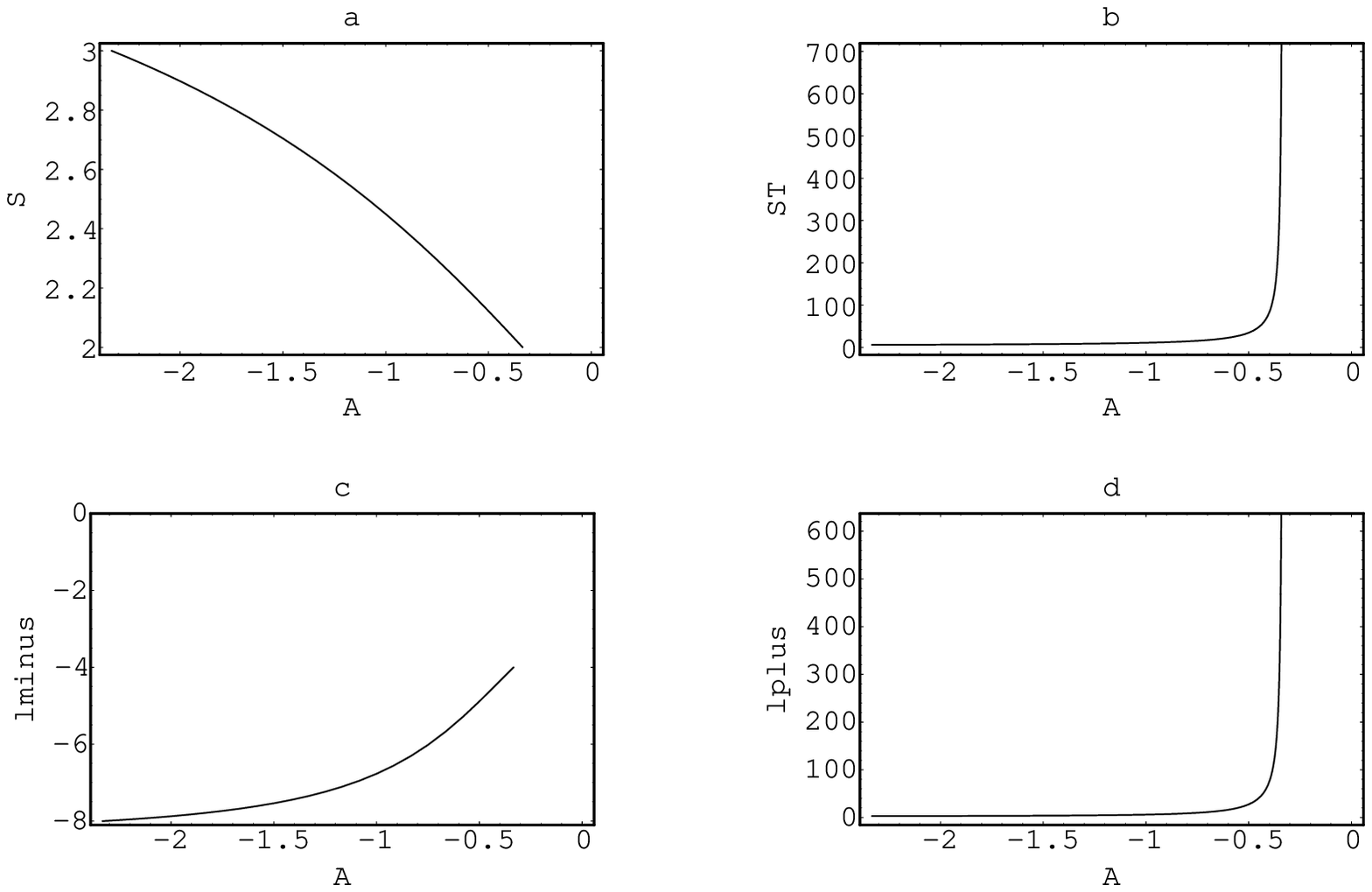}
\caption{The plots show the eigenvalues of the $\eta$-matrix for all
permissible values of $A$ ($-7/3$ to $-1/3$). From the top left,
going clockwise, we have $S$, $S+T$, $\lambda^+$ and $\lambda^-$.}
\label{fig:eig} }

It can also be seen that if we take, in particular, the seven moduli
case with all weights $a_i=1/3$ and take two of the fluxes to be
positive, (thus setting $A=-1$) we find the following spectrum of
$m^2/V_0$ : 10.90, 6.12, 2.45, -6.76. None of these are integer thus
providing a case of non-integral scalar masses in a dS
compactification.
\section{Applications to cosmology}
At first blush it appears that there are no particularly useful
applications to cosmology since in all cases the tachyonic
directions are rather steep (numerically they seem to be close to
the cases studied in \cite{Kallosh:2001gr}). Furthermore, since 2002
when such models were considered for dark energy, new data has
become available which point us closer towards a cosmological
constant (see, for example, \cite{Riess:2004}). This would mean that
we need even flatter potentials that those studied in
\cite{Kallosh:2001gr} if we would like to use the dS saddle points
to explain the current accelerating expansion of the universe.
However, it is worth taking a brief, closer look at the general
picture of this tachyonic ``fast roll inflation''
(\cite{Linde:2001}). We will follow the treatment in
\cite{Linde:2001} closely. Note that we are working in Planck units,
with the $4d$ Planck mass set equal to 1.

Near the unstable de Sitter point we can model a generic tachyonic
potential as:
\begin{equation}
V(\phi)=V_0-\frac{m^2\phi^2}{2}
\end{equation}
$V_0$ will of course be given by the value of the potential at the
saddle point and $m^2$ will depend on the value of the $a_i$, but
(as demonstrated above) will lie between $8V_0$ and $4V_0$. Defining
$\phi_{\ast}$ as the point at which the potential reaches half its
maximum value, we have:
\begin{equation}
\phi_{\ast}=\frac{\sqrt{V_0}}{m}
\end{equation}
The Hubble constant, $H$, remains fairly constant (at $H^2=V/3$)
while $\phi$ is in the range $\phi_0 < \phi < \phi_{\ast}$ and it can be
shown \cite{Linde:2001} that the total
expansion of the universe in this period is given by:
\begin{equation}
\frac{a(t_{\ast})}{a_0}\approx
e^{Ht_{\ast}}=\left(\frac{\phi_{\ast}}{\phi_0}\right)^{1/F}
\end{equation}
$F$ is given by:
\begin{equation}
F\left(\frac{m^2}{H^2}\right)=\sqrt{\frac{9}{4}+\frac{m^2}{H^2}}-\frac{3}{2}
\end{equation}

It is immediately clear that irrespective of $H$ and $m$ one could
achieve an arbitrary number of e-foldings by making $\phi_0$ small.
However, our ability to do this is constrained by the effects of
quantum fluctuations of the $\phi$ field. In \cite{Linde:2001} this
constraint is calculated to give a minimum value for $\phi_0$ of
$m/C$, where $C=O(10)$, which in turn means:
\begin{equation}
e^{Ht_{\ast}}\sim\left(\frac{10\phi_{\ast}}{m}\right)^{1/F}
\end{equation}
For our model $\phi_{\ast}\sim O(1)$ (in Planck units), so:
\begin{equation}
e^{Ht_{\ast}}\sim\left(\frac{10}{m}\right)^{1/F}
\end{equation}
Working with the steepest case gives $F^{-1}(8/3)=0.72$. To
calculate an estimate for the number of e-foldings we now need an
estimate for $m$.

Clearly, calculating $m$ is equivalent to calculating $V_0$ (since
$m^2=8V_0$), and we have the following expression for $V_0$ (from
\ref{potential}):
\begin{equation}
V=\frac{c_2^2}{48\pi V_X^3}\left(3+\sum_{j=1}^na_jh_j(h_j-3)\right)
\end{equation}
Recall that $V_X$ is given by \ref{volume}:
\begin{eqnarray}\label{volume}
V_X(s)=\prod_{k=1}^n\left(\frac{h_k}{\nu_k}\right)^{a_k}
=\prod_{k=1}^n\left(\frac{-c_2a_kh_k}{N_k}\right)^{a_k}
\end{eqnarray}
Continuing to look at the steepest case ($\eta=-8$) we can use our
earlier results (along with the observation that for this de Sitter
all the fluxes are negative and consequently $A=-7/3$) to get
$h_j=3$ and thus:
\begin{equation}
V_0=\frac{c_2^2}{16\pi V_X^3}\qquad\textrm{with},\qquad
V_X=\prod_{k=1}^n\left(\frac{-3c_2a_k}{N_k}\right)^{a_k}
\end{equation}
Since this is dependent on fluxes, it is manifestly tunable, and
thus we are (though with some limitations) free to chose $V_0$.
Before doing so, however, it is worth considering this tuning a
little more carefully.
One straightforwardly sees that to achieve a small $V_0$ we need to
make the volume, $V_X$, large. For volume we need the fluxes to be
as small as possible (i.e. $O(1)$ integers). Then, assuming the
$a_i$ are all of similar order we will have $V_X\sim c_2^{7/3}$ (and
similarly the compactification scale will go as $R\sim c_2^{1/3}$),
thus:
\begin{equation}
V_0\sim c_2^{-5}
\end{equation}
So, to obtain $V_0\sim10^{-120}$ we need $c_2\sim10^{24}$. Although
there is no known explicit construction of a $G_2$ manifold with
such a $c_2$, it does not appear that such a thing should be
prohibited. We note, though, that whilst these large volume
manifolds may exist they will only account for a small fraction of
the total number of $G_2$ flux compactifications \cite{acharya1}.

Now, $c_2\sim10^{24}$ implies that $V_X\sim10^{56}$ and that $R\sim
10^{8}$. As required, note that since there are still many orders of
magnitude difference between the size of the internal space and the
$dS$ space, we should still be within the regime where the
supergravity approximation is valid.
For a further consistency check we should try and find the
corresponding scale of Kaluza Klein excitations. To do this we first
need the $11d$ Planck mass. The calculation of $m_{kk}$ follows
\cite{acharya_modfix}.\footnote{Acharya obtains a very different
value  $m_{kk}$, since the chosen value of $c_2$ is much lower.
There the aim is to ensure that the fundamental Planck scale is the
same as the weak scale. Here, on the other hand, we are attempting
to address the cosmological constant problem, not the hierarchy
problem, hence the much larger value for $c_2$.} This is given, in
the usual fashion, by:
\begin{equation}
m_p^2\sim V_XM_p^2\sim c_2^{7/3}M_p^2\sim10^{56}M_p^2
\end{equation}
$m_p$ and $M_p$ are the $4d$ and $11d$ Planck masses respectively.
Then the KK scale will be given by:
\begin{equation}
m_{kk}\sim\frac{M_p}{R}\sim\left(\frac{1}{c_2}\right)^{\frac{1}{3}}M_p\sim\left(\frac{1}{c_2}\right)^{\frac{3}{2}}m_p\sim10^{-36}m_p\sim10^{-17}\textrm{GeV}
\end{equation}
While this is evidently extremely low compared to not only the
Planck scale, but also to any experimental bound, this alone is not
enough to mean that we should reject the model. As discussed in
\cite{acharya_modfix} the KK modes should transform under $7d$ gauge
transformations, whereas the standard model fields transform under
$4d$ gauge transformations. Accordingly, it would be difficult to
construct renormalizable interactions between the two and hence the
KK modes would be hard to detect.

Returning to the central calculation of this section, the final step
is to set $V_0$ equal to the observed value of the cosmological
constant ($10^{-120}M_p$). This in turn gives $m\sim10^{-60}$ and
thus:
\begin{equation}
e^{Ht_{\ast}}\sim\left(10^{61}\right)^{0.72}\sim10^{44}\sim e^{100}
\end{equation}
Of course, the universe can inflate $e^{100}$ times only if the
field $\phi$ initially was very close to the point $\phi = 0$.
However, to describe the present stage of expansion of the universe
we only need to have  one or two e-folds of accelerated expansion. A
numerical investigation of this issue shows that this can be
achieved if the initial value of the field $\phi$ is few times
smaller than the Planck mass. Such initial conditions seem quite
natural, especially if one takes into account that for $\phi$ a few
times greater than $M_{p}$ the universe rapidly collapses and cannot
support life as we know it.

This fact was the basis for an anthropic solution of the
cosmological constant problem in the models of a similar type in
\cite{Kallosh:2002gg}. All results obtained in \cite{Kallosh:2002gg}
should remain valid for our class of models, up to coefficients
$O(1)$. But in our case in addition to the anthropic constraints
$V_{0}\lesssim10^{-120}$ and $\phi \lesssim M_{p}$  we also have a
related anthropic constraint $c_2 \gtrsim10^{24}$. It remains to be
seen whether the models with such enormously large value of $c_{2}$
exist, or if this requirement rules out the class of models proposed
in \cite{acharya1}.

Furthermore, the combination of this requirement to have a large
$c_2$, the relatively sensitive initial conditions and the absence
of light KK-Standard Model interactions begins to have the
appearance of a tower of ``ifs''. While this does not preclude us
from considering this class of models for cosmology, it does make
them somewhat less than desirable.

\section{Conclusions}

We have explored the consequences of the models of \cite{acharya1}
with specific reference to the kind of cosmologies they can give
rise to. While it appears that such models do not allow slow roll
regimes, it may be the case that the tachyonic directions of the
moduli potential can give rise to fast roll inflation, which in turn
can provide a mechanism for the current accelerated expansion of the
universe. However, in order to achieve the kind of acceleration seen
today we are forced to only consider manifolds where the topological
invariant $c_2$ takes large values. This may seem somewhat
unnatural, but there appears to be no a priori reason that such
manifolds should not exist and could not be constructed.

As a final point, it is worth noting that whilst this set of models
does not appear to give rise to the full tapestry of cosmologies
that may describe our universe, that does not necessarily deprive
the analysis of value. The landscape of vacua is vast and so the
possibility of excluding sections of it as unusable is, in some
ways, as useful as adding new models.

\newpage

\leftline{\bf Acknowledgments}

We are  grateful to B.S.~Acharya, F.~Denef, A.~Linde, and S.~Kachru
for valuable discussions. This work is supported by NSF grant
0244728. N.S. is also supported the U.S. Department of Energy under
contract number DE-AC02-76SF00515.

\end{document}